# Light Curve Analysis of Ground-Based Data
# from Exoplanets Transit Database

**F. Davoudi[1,2]; S. J. Jafarzadeh[1,2]; A. Poro[1,2]; O. Basturk[3]; S. Mesforoush[2]; A. Fasihi Harandi[2];**
**MJ. Gozarandi[1,2]; Z. Zare Mehrjardi[1,2]; P. D. Maley[4]; S. Khakpash[2,5]; K. Rokni[2]; A. Sarostad[2]**

[1] The International Occultation Timing Association-Middle East Section, info@iota-me.com
[2] The Six Summer School of Astronomy (IOTA/ME), Khayyam Observatory, Mahdasht, Iran
[3] Ankara University, Faculty of Science, Astronomy and Space Sciences Department, TR-06100, Tandogan, Ankara, Turkey
[4] The International Occultation Timing Association, Carefree AZ, USA
[5] Department of Physics, Lehigh University, 16 Memorial Drive East, Bethlehem, PA 18015, USA

**Abstract**

Photometric observations of exoplanet transits can be used to derive the orbital and physical parameters of an exoplanet. We analyzed several transit light curves of exoplanets that are suitable for ground-based observations whose complete information is available on the Exoplanet Transit Database (ETD). We analyzed transit data of planets including HAT-P-8 b, HAT-P-16 b, HAT-P-21 b, HAT-P-22 b, HAT-P-28 b and HAT-P-30 b using the AstroImageJ (AIJ) software package. In this paper, we investigated 82 transit light curves from ETD, deriving their physical parameters as well as computing their mid-transit times for future Transit Timing Variation (TTV) analyses. The Precise values of the parameters show that using AIJ as a fitting tool for follow-up observations can lead to results comparable to the values at the NASA Exoplanet Archive (the NEA). Such information will be invaluable considering the numbers of future discoveries from ground and space-based exoplanet surveys.

## Introduction

Ground-based observations alongside space missions have great scientific potential in the discovery of new exoplanets and validating their parameters. Various physical parameters can be deduced using transit light curves. Of course, the number and accuracy of these parameters depend on the accuracy of the photometry, the type of exoplanet system, and the initial assumptions of the orbit using Kepler's laws. Ground-based exoplanet surveys have made a lot of contributions to this field. In this paper, we chose cases derived from Hungarian-made Automated Telescope (HAT) surveys.

The Hungarian-made Automated Telescope Network (HATNet) has been in operation since 2003, with the key science goal being the discovery and accurate characterization of Transiting Extrasolar Planets (TEPs) around bright stars in the Northern hemisphere (Bakos et al. 2004). HATNet consists of six wide field automated telescopes: four of these are located at the Fred Lawrence Whipple Observatory (FLWO) in Arizona and two on the roof of the Submillimeter Array hangar of the Smithsonian Astrophysical Observatory (SAO) in Hawaii. Furthermore, the HATSouth network was the first global network of telescopes using identical instrumentation. They have been important contributors to the rapidly developing field of exoplanets, motivating and influencing observational techniques, theoretical studies, and also actively shaping future instrumentation for the detection and characterization of exoplanets (Bakos, 2018).

Throughout this article, we improved the parameters of the planets in proportion to the Exoplanet Transit Database (ETD)[1]. We analyzed several transit curves of exoplanets based on data from ETD. ETD is a project of the Variable Star and Exoplanet Section of the Czech Astronomical Society. It is created to supply observers with such useful information as transit predictions, Transit Timing Variation (TTV); in addition, it enables the user to plot light curves depicting their depth and duration. The main goal of the ETD is to gather all available light curves from professional and also amateur astronomers. They use a simple analytical model of the transit to calculate the central time of transit, its duration and the depth of the transit. These values are then plotted into the observed computed diagrams (O-C) that represent the last part of the application (Poddan'y et al. 2009). In the Data section below, we discuss how to select data from the ETD and specifications of the observations.

In the Method section below, we describe the factors that show the distinction between AIJ and ETD discussing how to use the AIJ package efficiently. AIJ is a graphical software package for general image processing. It simplifies light curve plotting, especially for applications requiring ultra-precise light curves such as exoplanet transits (Mandel & Agol, 2002). AIJ is a Java program and its astronomical algorithms are based on codes from JSkyCal; this package is compatible with all operating systems (Collins et al. 2017). Transit parameters are calculated from the best fit model by AIJ including the planetary radius in units of the stellar radius, $R_P / R_*$, the orbital inclination $i$, the transit duration $t_{14}$ and the mid-transit time $t_c$. In the Data Analysis section, we summarize the calculated values of the parameters of each exoplanet using AIJ and determine uncertainties in the parameters of transit with a differential evolution Markov Chain Monte Carlo (MCMC) simulation. Finally, in the Results section, we compared values of the output parameters of planets with that from the NASA Exoplanet Archive (the NEA) database[2].

The NEA is operated by the California Institute of Technology under contract with the National Aeronautics and Space Administration's Exoplanet Exploration Program. This archive combines a database of the confirmed exoplanets and host star properties with key public data sets from space and ground-based surveys; it also provides quantitative analysis tools to work with this data. Examples of the data included are stellar parameters (positions, magnitudes, and temperatures), exoplanet parameters (masses and orbital parameters), and discovery/characterization data (published radial velocity curves, photometric light curves, images, and spectra). The contents of each data set are fully described as part of the archive documentation. The Exoplanet Archive also includes over 2.9 million light curves, including public data from space missions and several ground-based surveys (Akesoni et al. 2013).
So, it is possible to derive valuable information by analyzing transit light curves of exoplanets, provided by even amateur astronomers with access to relatively modern observing equipment, by using off-the-shelf analysis software as AIJ and web services such as EXOFAST. The mid-transit times as the results of our work will constitute a processed input to perform a Transit Timing Variation (TTV) analysis for these exoplanets.

## Data Sets

Exoplanets that we have studied in this paper are HAT-P-8 b (D.W. Latham et al. 2009), HAT-P-16 b (L.A. Buchhave et al. 2010), HAT-P-21 b (G.Á. Bakos et al. 2011), HAT-P-22 b (G.Á. Bakos et al. 2011), HAT-P-28 b (L.A. Buchhave et al. 2011), and HAT-P-30 b (J.A. Johnson et al. 2011) whose light curves are taken from the ETD. We chose these planets because the number of light curves of their transits is appropriate and their observations have been made in the last ten years. They have suitable brightnesses to enable ground-based observations (maximum magnitude 15). In addition to the suitability of data quality, we made a selection by eye. For example, we chose those for analysis that have sufficiently out-of-transit data points and show a clear decrease of the brightness in the light curve during a transit. The orbital period of each exoplanet is less than 5 days. Duration of transits is less than 0.2 day and these planets only orbit around one star.
Data are provided by different observers from different geographical locations. For example, for HAT-P-8 b 14 light curves were examined for which their observational information is listed in Table.2. The first column refers to the data file number, the next five columns refer to the observer's name, observation dates, the filter type, optical size, and the CCD model used, respectively. The last column refers to the data quality of individual light curves which is indicated by a number from 1 (for the best) to 5 (for the worst) based on the records in the ETD that are divided into five groups according to their data quality index DQ. While computing the DQ index of the light curve, the following relation is used

$$\alpha = \frac{\delta}{S}\sqrt{\rho} \qquad (1)$$

Here α is a temporary data quality index, S is the mean absolute deviation of the data from their fit and $\rho = N/l$ is the data sampling, where $l$ is the length of observing run in minutes. The number $\alpha$ is further transformed for better lucidity to the scale from 1 to 5 where 1 presents the best quality data and the value 5 the worst data (Table 1). They provide only the results of mid-transit time, transit depth and length of the transit, not the whole light curve (Poddan'y et al. 2009).

**Table 1. The distribution of the quality of the light curves according to their DQ index**

| DQ index | 1 | 2 | 3 | 4 | 5 |
|----------|---|---|---|---|---|
| Threshold | $\alpha \geq 9.5$ | $9.5 > \alpha \geq 6.0$ | $6.0 > \alpha \geq 2.5$ | $2.5 > \alpha \geq 1.3$ | $1.3 > \alpha$ |

In the following tables data for the six exoplanets light curve information are shown. Tables 2-7 illustrate initial data information for HAT-P-8 b, HAT-P-16 b, HAT-P-21 b, HAT-P-22 b, HAT-P-28 b and HAT-P-30 b respectively.

**Table 2. Initial data information for HAT-P-8 b**

| NO. | Observer | Observation Date | Filter | Optic size (mm) | CCD Model | DQ |
|-----|----------|------------------|--------|-----------------|-----------|-----|
| 1 | H. Kučáková | 2009-07-22 | R | 200 | ST-8XME | 3 |
| 2 | B. Norby | 2009-08-03 | R | 254 | - [3] | 3 |
| 3 | A. Ayiomamitis | 2009-08-27 | Clear | 162.56 | ST-10XME | 3 |
| 4 | G. Srdoc | 2009-08-27 | R | 304.8 | - | 3 |
| 5 | H. Kučáková | 2009-08-29 | R | 200 | ST8-XME | 3 |
| 6 | J. Trnka | 2009-08-29 | Clear | 200 | ST-9E | 3 |
| 7 | G. Srdoc | 2009-08-30 | R | 304.8 | - | 3 |
| 8 | F. Hormuth | 2010-09-02 | R | 1230 | DLR-MKIII | 1 |
| 9 | N. Ruocco | 2010-09-02 | R | 304.8 | ST-7 | 3 |
| 10 | G. Marino | 2010-09-02 | Clear | 317.5 | QSI 516wsg | 3 |
| 11 | K. Hose | 2011-03-17 | R | 200 | ST-8 | 3 |
| 12 | *R. Uhlar* | 2011-08-30 | R | 304.8 | ST9-XE | 5 |
| 13 | K. Hose | 2012-09-23 | R | 200 | - | 2 |
| 14 | *R. Naves* | 2016-08-27 | R | 304.8 | ST10XME | 2 |

**Table 3. Initial data information for HAT-P-16 b**

| NO. | Observer | Observation Date | Filter | Optic size (mm) | CCD Model | DQ |
|-----|----------|------------------|--------|-----------------|-----------|-----|
| 1 | M. Vrašťák | 2010-09-30 | Clear | 240 | G2 | 2 |
| 2 | J. Világi et al. | 2010-11-10 | Clear | 600 | - | 2 |
| 3 | S. Shadick | 2010-11-01 | Clear | 304.8 | STL1301 | 3 |
| 4 | S. Shadick | 2011-10-17 | Clear | 304.8 | ST10XME | 2 |
| 5 | J. Trnka | 2011-10-01 | Clear | 200 | ST-9E | 3 |
| 6 | T. Sauer | 2011-10-06 | R | 203.2 | ST-10 | 3 |
| 7 | S. Shadick | 2011-10-17 | I | 304.8 | STL1301 | 3 |

---

[3] When the information about the CCD is not available, it is represented by a dash.

| 8 | R. Garcia | 2012-01-10 | V | 355.6 | - | 3 |
| 9 | J. Mravik et al. | 2012-10-23 | Clear | 203.2 | ST-7 | 3 |
| 10 | F.G. Horta | 2013-09-12 | V | 305 | FLI PL1001E-1 | 3 |
| 11 | P. Benni | 2013-11-06 | Clear | 203.2 | ST8300M | 3 |
| 12 | A. Ayiomamitis | 2013-11-18 | Clear | 305 | ST-10XME | 3 |
| 13 | F. Scaggiante et al. | 2014-01-06 | R | 410 | KAF402ME | 3 |
| 14 | D.Molina | 2015-11-09 | Clear | 2000 | ST-8XME | 2 |
| 15 | F. Campos | 2015-11-23 | R | 200 | ST-8XME | 3 |
| 16 | D. Molina | 2016-12-27 | Clear | 203 | ST-8XME | 2 |

**Table 4. Initial data information for HAT-P-21 b**

| NO. | Observer | Observation Date | Filter | Optic size (mm) | CCD Model | DQ |
|---|---|---|---|---|---|---|
| 1 | R. Dreven | 2011-03-09 | R | 200 | G1 300 | 3 |
| 2 | R. Zambelli | 2011-04-14 | R | 254 | ST-8XME | 2 |
| 3 | L. Trevmosa | 2012-01-04 | Clear | 132 | - | 3 |
| 4 | P. Benni | 2014-02-14 | Clear | 203.2 | ST8300M | 3 |
| 5 | J.P. Jacobsen | 2016-05-19 | Clear | 355.6 | - | 3 |

**Table 5. Initial data information for HAT-P-22 b**

| NO. | Observer | Observation Date | Filter | Optic size (mm) | CCD Model | DQ |
|---|---|---|---|---|---|---|
| 1 | L. Brát | 2011-02-24 | Clear | 200 | G1-0300 | 3 |
| 2 | R. Zambelli | 2011-03-14 | R | 254 | ST-8XME | 2 |
| 3 | G. Marino | 2012-01-26 | R | 200 | ST-7 | 3 |
| 4 | J.L. Haro | 2013-02-11 | V | 250 | ST-7XME | 3 |
| 5 | A. Ayiomamitis | 2013-03-14 | Clear | 305 | ST-10XME | 2 |
| 6 | V.P. Hentunen | 2014-02-06 | R | 355.6 | ST-8XME | 3 |
| 7 | M. Salisbury | 2014-05-07 | R | 400 | ST-10XME | 2 |
| 8 | CAAT group | 2015-08-10 | Clear | 406.4 | ST-7XME | 3 |
| 9 | R.N. Nogues | 2015-08-10 | R | 305 | ST-8XME | 3 |
| 10 | V.P. Hentunen | 2017-01-12 | R | 355.6 | ST-8XME | 2 |
| 11 | D. Molina | 2018-12-11 | Clear | 203 | ST-8XME | 2 |

**Table 6. Initial data information for HAT-P-28 b**

| NO. | Observer | Observation Date | Filter | Optic size (mm) | CCD Model | DQ |
|---|---|---|---|---|---|---|
| 1 | H. Kučáková | 2011-09-05 | Clear | 304.8 | ST-8XME | 5 |
| 2 | F. Lomoz | 2011-09-05 | Clear | 254 | G2-8300 | 3 |
| 3 | M. Zíbar | 2011-09-05 | Clear | 203.2 | Atic 16 IC | 3 |
| 4 | L. Brát | 2011-09-05 | Clear | 200 | G1-0300 | 3 |
| 5 | J. Trnka | 2011-10-01 | Clear | 300 | G1-0300 | 3 |
| 6 | L. Brát | 2011-10-01 | Clear | 200 | G1-0300 | 3 |
| 7 | F. Lomoz | 2011-10-01 | Clear | 300 | ST2000XM | 4 |
| 8 | J. Trnka | 2011-10-24 | Clear | 200 | ST-9E | 3 |
| 9 | S. Shadick | 2011-12-27 | Clear | 304.8 | ST-10XME | 3 |
| 10 | P. Benni | 2013-08-09 | Clear | 279.4 | ST-8XME | 3 |
| 11 | P. Benni | 2013-11-21 | Clear | 279.4 | ST-8XME | 3 |
| 12 | P. Benni | 2014-02-06 | Clear | 279.4 | ST-8XME | 3 |
| 13 | J. Gonzalez | 2014-08-20 | Clear | 235 | ST-8XME | 3 |

| 14 | J. Garlitz | 2014-10-22 | Clear | 304.8 | SBIG402 | 3 |
| 15 | M. Bretton | 2015-07-22 | Clear | 430 | STL-11K | 2 |
| 16 | M. Bretton | 2015-11-02 | Clear | 254 | T820 | 2 |
| 17 | M. Fleenor | 2016-12-27 | Clear | 355 | ST-10XME | 2 |
| 18 | C. Rude et al. | 2017-01-12 | R | 500 | SBIG6303E | 3 |
| 19 | M. Bretton | 2018-12-11 | Clear | 820 | FLI PL230 | 1 |
| 20 | Y. Jongen | 2018-12-29 | Clear | 355.6 | STLX11002 | 3 |

**Table 7. Initial data information for HAT-P-30 b**

| NO. | Observer | Observation Date | Filter | Optic size (mm) | CCD Model | DQ |
|---|---|---|---|---|---|---|
| 1 | R. Uhlař | 2011-03-30 | R | 200 | - | 3 |
| 2 | S. Shadick | 2012-04-25 | Clear | 254 | ST-10XME | 3 |
| 3 | J. Gonzalez | 2012-01-26 | R | 235 | ST-8XME | 3 |
| 4 | S. Shadick | 2012-04-02 | I | 254 | ST-10XME | 3 |
| 5 | J. Gonzalez | 2012-02-21 | Clear | 235 | ST-8XME | 3 |
| 6 | G. Marino | 2012-02-21 | R | 200 | ST-7 | 3 |
| 7 | S. Shadick | 2012-03-19 | R | 355.6 | ST-10XME | 3 |
| 8 | R. Naves | 2012-11-15 | R | 305 | ST-8XME | 3 |
| 9 | S. Shadick | 2013-01-18 | Clear | 254 | ST-10XME | 3 |
| 10 | V. P. Hentunen | 2014-02-06 | Clear | 355.6 | ST-8XME | 3 |
| 11 | G. Marino | 2014-02-18 | I | 250 | ST-7 | 3 |
| 12 | E. Sokov et al. | 2014-05-07 | V | 500 | ST-L 1001E | 3 |
| 13 | F. Scaggiante et al. | 2015-07-22 | R | 410 | KAF402ME | 3 |
| 14 | A. Christophe et al. | 2015-08-10 | R | 600 | STL6303E | 3 |
| 15 | D. Molina | 2018-12-11 | Clear | 203 | ST-8XME | 3 |
| 16 | F. Campos | 2019-02-18 | R | 203.2 | ST-8XME | 2 |

All observations have been performed by using CCD detectors attached on small-to-medium size telescopes. The smallest telescope has a primary with the diameter 132 mm, while the largest has 2000 mm-diameter mirror. The average optics size in all observations is 323.34 mm (12.73 inches). Moreover, the majority of the telescopes were in the range of 200-400 mm (Table 8).

**Table 8. The size of the optics used in the observations**

| Planet | The smallest optic | The largest optic | Average optics size |
|---|---|---|---|
| HAT-P-8 b | 162.56 mm | 1230 mm | 320.57 mm |
| HAT-P-16 b | 200 mm | 2000 mm | 396.41 mm |
| HAT-P-21 b | 132 mm | 355.6 mm | 228.96 mm |
| HAT-P-22 b | 200 mm | 406.4 mm | 294.05 mm |
| HAT-P-28 b | 200 mm | 820 mm | 317. 97 mm |
| HAT-P-30 b | 200 mm | 600 mm | 300.90 mm |

**Method**

Photometric datasets from ETD include differential magnitudes with respect to the mid-time of CCD exposures, in either Julian Dates (JD) or Heliocentric Julian Dates (HJD). Because of the high accuracy of Barycentric Julian Dates (BJD) in comparison to other reference frames, the HJD was formally deprecated by the International Astronomical

Union (IAU) in 1991 in favor of the BJD, a time referenced to the Solar System Barycenter (SSB) (Eastman et al. 2010). Therefore we converted our times from JD_UTC or HJD_UTC to BJD_TDB through the web service[4].

For each exoplanet, we processed its data with the AIJ. A transiting exoplanet is modeled as an eclipse of a spherical star by an opaque planetary sphere. The model is characterized by six physical values, plus a baseline flux level, $F_0$. The six physical parameters are the planetary radius in units of the stellar radius, $R_P / R_*$, the distance between the planet and star scaled in units of the stellar radius, $a / R_*$, the transit centre time, $T_C$, the impact parameter of the transit, $b$, and the quadratic limb darkening parameters, $u_1$ and $u_2$. The orbital inclination can be calculated from the model parameters as

$$i = \cos^{-1}\left(b\frac{R_*}{a}\right). \qquad (2)$$

The best fit model is found by minimizing $\chi^2$ of the model residuals using the downhill simplex method to find the local minima. Light curve detrending is accomplished by including a $\chi^2$ contribution for each selected detrend parameter in the overall light curve fit. The $\chi^2$ contribution for all n detrend parameters are calculated at each step of the fitting process as

$$x_D^2 = \sum_{k=1}^{m} \frac{(o_k - (\sum_{j=1}^{n} c_j D_{jk}) - E_k)^2}{\sigma_k^2} \qquad (3)$$

here j indexes the detrend parameters, k indexes the samples of the light curve, m is the total number of samples in the light curve, $o_k$ is the observed normalized differential target flux, $c_j$ is the fitted linear coefficient for the detrend parameter values $D_{jk}$, $E_k$ is the expected value of the flux (which is the normalized transit model value corresponding to the time of the $k_{th}$ data sample), and $\sigma_k$ is the error in the normalized differential target flux for each sample.

Furthermore, the best linear fit of a trend dataset (airmass in this case) to the light curve is found by using AIJ. It is possible to see how the fit is improved by detrending for the airmass. A good guide for whether a detrending parameter is effective is to view the Bayesian Information Criterion (BIC) value (P.M. Bentler 1995). If it reduces by more than 2.0, then airmass should be detrended. In other words, the fit with the lowest BIC value indicates the preferred model. The BIC is calculated as

$$BIC = \chi^2 + p \ln n \qquad (4)$$

Where p is the number of fitted parameters, and n is the number of fitted data points. The $\chi^2$ contribution represents the goodness of the linear fit of the detrend parameter to the light curve after subtracting the light curve model corresponding to the current fit step (Collins et al. 2017).

In AIJ the period of the exoplanet's orbit (P) is not well constrained by the transit model, but its value will impact the best fit values of some of the fitted parameters; therefore P should be manually entered by the user. Moreover, the host star parameters are entered to estimate the transit parameters $\left(\frac{R_P}{R_*}, \frac{a}{R_*}, b\right)$. The host star radius can be retrieved from the NEA and the $u_1$ and $u_2$ are the quadratic limb darkening coefficients, which can be found by interpolation based on the stellar parameters and the passband of the observation using the limb darkening coefficient calculator for all the frequently used passbands other than the clear observations[5]. Otherwise, entering any other parameter of the host star will produce a rough estimate of $R_*$ based on the zero age main sequence assumption (Collins et al., 2017).

Our analysis with the AIJ is more thorough in that it is based on a physical model while The ETD assumes that the planet and the host star have dark and limb-darkened disks with a radius ratio of k = $R_P / R_*$ keeping in mind that the planetary radius is much smaller than the radius of the host star, k $\leq$ 0.2. The ETD uses the Levenberg-Marquardt non-linear least squares fitting algorithm (Press et al. 2007). The algorithm requires the initial values of the parameters and partial derivatives of the fitted function. For a limb-darkened star, the depth of the transit $\delta$ is

---

determined by radius ratio k, impact factor b, and limb darkening coefficient $c_1$. The ETD keeps $c_1$ fixed at an arbitrary value $c_1 = 0.5$ in all cases because the effect on other parameters is rather negligible, usually smaller than the error bars. The export value of the depth is then evaluated as

$$\delta = -2.5 \log [\min z\, F\, (z, k, c_1)] \qquad (5)$$

z is a function with four parameters $t_i$, $T_c$, $t_{14}$, and b, where b is the impact parameter (Poddan'y et al. 2009). For these reasons, we didn't use the analysis of data by ETD.

As an example of what we did in this work, the data processing with AIJ for the exoplanet HAT-P-30 b is shown in Figure 1. The raw normalized light curve and the light curve after fitting the exoplanet transit model to the data shows how the airmass detrending parameter was used to improve the fit.

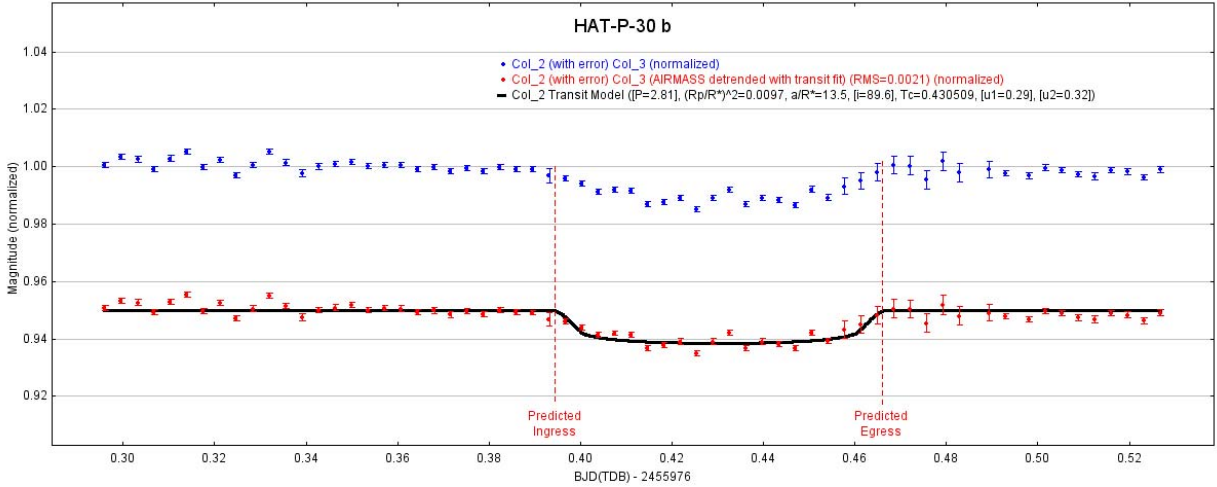

**Figure 1. Multi-plot example for HAT-P-30 b transit using the airmass detrend parameter. The raw data are shown as solid blue dots which are plotted with no fit, while the solid red dots show the same data assuming that it will reflect a dip in brightness due to the exoplanet's transit, which is shown by the black line through the data.**

**Data Analysis**

In this section, we summarize the calculation of the parameter values for each exoplanet using AIJ. Since AIJ does not find the error for each of those model parameters, uncertainties in the parameters are determined with a differential evolution MCMC simulation through a free web-based tool (EXOFAST). The EXOFAST takes flux and time (which are output of the AIJ) as input and provides parameters values in addition to the errors of some of them. This method is a standard tool in exoplanet research when the goal is to decrease the overall analysis error. Therefore, we should find the lowest $\chi^2$ to find this minimum:

$$\chi^2 = \sum \left( \frac{D_i - M_i}{\sigma_i} \right)^2 \qquad (6)$$

in which a given data set D is described by a given model M. The online EXOFAST tool is based on this method (J. Eastman et al. 2013) and has been used to derive the uncertainties only by fixing the parameter values found in the analyses[6]. With EXOFAST a set of parameters is selected and the $\chi^2$ is evaluated for that set. Then a different set of parameters is randomly selected, the $\chi^2$ is calculated and compared to that of the previous set to find the best set of parameters values. To do this, EXOFAST needs three columns of data including BJD_TDB, flux, flux uncertainty. Since AIJ converts magnitude to the normalized relative flux, we can derive the uncertainties in certain parameters with the help of this tool.

---

[6] http://astroutils.astronomy.ohiostate.edu/exofast/example.html

We listed the uncertainties of all the parameters in Tables 9-14. The tables contain the parameters of different data for every single exoplanet. Each row in these tables shows the calculated parameters for each light curve mentioned in the Data section (For example Table 9 shows all parameters calculated from 14 light curves in the HAT-P-8 b dataset). The first column is the number for each light curve of the planet; the second column ($R_P$) is the radius of planet in terms of the radius of Jupiter; the third column is $k = \frac{R_P}{R_*}$ which shows the ratio of the planet radius to the stellar radius; the fourth column ($t_{14}$) is the total transit time from the first contact to the fourth contact; the fifth column ($i$) is the orbital inclination (deg); the sixth column (BIC) is the BIC value. The seventh column is the mid-transit times (T$_c$) and other two columns are the errors FWHM duration ($\sigma_{T_{FWHM}}$), and the transit depth ($\sigma_{depth}$). We calculated the error propagation of k and $R_P$ from the following relations:

$$\sigma_k = \frac{\sigma_{depth}}{2k}, \qquad \sigma_{R_p} = k\sigma_{R_*} + R_*\sigma_k \qquad (7)$$

**Table 9. Parameters calculated for HAT-P-8 b**

| NO. | $R_P$(R$_J$) | k | T$_{14}$(day) | $i$(°) | BIC | $T_c$ (BJD$_{TDB}$) | $\sigma_{T_{FWHM}}$ | $\sigma_{depth}$ |
|---|---|---|---|---|---|---|---|---|
| 1 | 1.68 ± 0.10 | 0.109 ± 0.001 | 0.168 ± 0.002 | 90.0 | 60.31 | 2455034.4913 ± 0.0002 | 0.0018 | 0.00022 |
| 2 | 1.45 ± 0.10 | 0.094 ± 0.001 | 0.163 ± 0.002 | 88.6 | 60,00 | 2455046.7891 ± 0.0007 | 0.0015 | 0.00020 |
| 3 | 1.54 ± 0.11 | 0.100 ± 0.001 | 0.197 ± 0.003 | 89.9 | 56.06 | 2455071.3841 ± 0.0010 | 0.0021 | 0.00029 |
| 4 | 1.43 ± 0.13 | 0.094 ± 0.003 | 0.173 ± 0.008 | 88.8 | 52.42 | 2455071.4027 ± 0.0024 | 0.0049 | 0.00062 |
| 5 | 1.47 ± 0.11 | 0.094 ± 0.001 | 0.171 ± 0.004 | 88.2 | 56.50 | 2455071.4112 ± 0.0012 | 0.0024 | 0.00034 |
| 6 | 1.64 ± 0.12 | 0.104 ± 0.002 | 0.187 ± 0.007 | 87.6 | 56.74 | 2455071.4123 ± 0.0021 | 0.0043 | 0.00061 |
| 7 | 1.80 ± 0.11 | 0.118 ± 0.001 | 0.169 ± 0.004 | 86.0 | 52.93 | 2455074.4787 ± 0.0012 | 0.0024 | 0.00044 |
| 8 | 1.46 ± 0.10 | 0.094 ± 0.000 | 0.169 ± 0.001 | 88.3 | 51.70 | 2455437.4888 ± 0.0003 | 0.00072 | 0.00010 |
| 9 | 1.75 ± 0.11 | 0.114 ± 0.001 | 0.156 ± 0.004 | 88.0 | 48.44 | 2455437.4897 ± 0.0012 | 0.0024 | 0.00042 |
| 10 | 1.46 ± 0.11 | 0.094 ± 0.001 | 0.166 ± 0.004 | 88.6 | 41.90 | 2455440.5651 ± 0.0012 | 0.0024 | 0.00036 |
| 11 | 1.39 ± 0.10 | 0.089 ± 0.001 | 0.176 ± 0.002 | 88.4 | 50.56 | 2455452.8620 ± 0.0008 | 0.0016 | 0.00020 |
| 12 | 1.47 ± 0.10 | 0.094 ± 0.001 | 0.166 ± 0.002 | 88.3 | 42.19 | 2455800.4924 ± 0.0007 | 0.0015 | 0.00024 |
| 13 | 1.43 ± 0.10 | 0.094 ± 0.000 | 0.167 ± 0.001 | 88.2 | 50.89 | 2456175.8100 ± 0.0005 | 0.0010 | 0.00014 |
| 14 | 1.22 ± 0.11 | 0.077 ± 0.001 | 0.150 ± 0.001 | 88.5 | 51.06 | 2457615.5360 ± 0.0011 | 0.0021 | 0.00028 |

**Table 10. Parameters calculated for HAT-P-16 b**

| NO. | $R_P$(R$_J$) | k | T$_{14}$(day) | $i$(°) | BIC | $T_c$ (BJD$_{TDB}$) | $\sigma_{T_{FWHM}}$ | $\sigma_{depth}$ |
|---|---|---|---|---|---|---|---|---|
| 1 | 1.19 ± 0.05 | 0.099 ± 0.001 | 0.1287 ± 0.0021 | 88.4 | 45.75 | 2455463.4190 ± 0.0006 | 0.0012 | 0.00026 |
| 2 | 1.46 ± 0.05 | 0.121 ± 0.001 | 0.1263 ± 0.0003 | 88.5 | 66.40 | 2455463.4204 ± 0.0001 | 0.00021 | 0.00029 |
| 3 | 1.29 ± 0.05 | 0.107 ± 0.001 | 0.1231 ± 0.0004 | 88.0 | 61.78 | 2455482.8513 ± 0.0001 | 0.00024 | 0.00028 |
| 4 | 1.34 ± 0.05 | 0.112 ± 0.000 | 0.1300 ± 0.0002 | 88.0 | 70.64 | 2455829.8521 ± 0.0002 | 0.00012 | 0.00016 |
| 5 | 1.20 ± 0.06 | 0.101 ± 0.002 | 0.1245 ± 0.0024 | 88.6 | 53.79 | 2455835.3994 ± 0.0007 | 0.0014 | 0.00045 |
| 6 | 1.34 ± 0.06 | 0.112 ± 0.001 | 0.1236 ± 0.0005 | 88.3 | 55.34 | 2455835.4019 ± 0.0001 | 0.00034 | 0.00042 |
| 7 | 1.29 ± 0.05 | 0.107 ± 0.001 | 0.1270 ± 0.0001 | 89.0 | 65.40 | 2455843.7303 ± 0.0003 | 0.00022 | 0.00030 |
| 8 | 1.19 ± 0.06 | 0.099 ± 0.001 | 0.1312 ± 0.0004 | 89.1 | 52.10 | 2455935.3303 ± 0.0001 | 0.00024 | 0.00033 |
| 9 | 1.58 ± 0.23 | 0.131 ± 0.024 | 0.1442 ± 0.0071 | 87.4 | 50.33 | 2456218.4696 ± 0.0020 | 0.00414 | 0.00633 |
| 10 | 1.55 ± 0.05 | 0.129 ± 0.001 | 0.1231 ± 0.0006 | 88.0 | 57.83 | 2456540.4996 ± 0.0002 | 0.00040 | 0.00047 |
| 11 | 1.33 ± 0.05 | 0.111 ± 0.001 | 0.1251 ± 0.0002 | 88.5 | 65.93 | 2456601.5685 ± 0.0002 | 0.00016 | 0.00022 |
| 12 | 1.26 ± 0.06 | 0.105 ± 0.001 | 0.1263 ± 0.0006 | 89.0 | 60.69 | 2456604.3436 ± 0.0001 | 0.00036 | 0.00041 |
| 13 | 1.39 ± 0.05 | 0.116 ± 0.001 | 0.1211 ± 0.0009 | 89.0 | 62.18 | 2456629.3272 ± 0.0001 | 0.00055 | 0.00041 |
| 14 | 1.49 ± 0.05 | 0.124 ± 0.001 | 0.1302 ± 0.0005 | 88.7 | 53.96 | 2457334.4231 ± 0.0001 | 0.00029 | 0.00038 |
| 15 | 1.30 ± 0.05 | 0.108 ± 0.002 | 0.1257 ± 0.0004 | 89.9 | 49.25 | 2457345.5264 ± 0.0001 | 0.00025 | 0.00043 |
| 16 | 1.21 ± 0.05 | 0.100 ± 0.001 | 0.1248 ± 0.0001 | 88.4 | 50.77 | 2457706.4042 ± 0.0001 | 0.00023 | 0.00023 |

**Table 11. Parameters calculated for HAT-P-21 b**

| NO. | $R_P$(R$_J$) | K | T$_{14}$(day) | $i$(°) | BIC | $T_c$ (BJD$_{TDB}$) | $\sigma_{T_{FWHM}}$ | $\sigma_{depth}$ |
|---|---|---|---|---|---|---|---|---|
| 1 | 1.08 ± 0.18 | 0.100 ± 0.002 | 0.1532 ± 0.0010 | 88.6 | 56.04 | 2455627.4771 ± 0.0002 | 0.00059 | 0.00053 |
| 2 | 1.10 ± 0.17 | 0.102 ± 0.001 | 0.1701 ± 0.0006 | 88.5 | 50.09 | 2455660.4595 ± 0.0001 | 0.00037 | 0.00034 |
| 3 | 1.04 ± 0.18 | 0.096 ± 0.002 | 0.1504 ± 0.0007 | 88.8 | 52.29 | 2455928.5470 ± 0.0002 | 0.00042 | 0.00045 |
| 4 | 1.09 ± 0.17 | 0.102 ± 0.001 | 0.1531 ± 0.0004 | 88.5 | 67.15 | 2456699.8261 ± 0.0001 | 0.00025 | 0.00028 |

| 5 | 0.98 ± 0.19 | 0.091 ± 0.002 | 0.1438 ± 0.0010 | 88.7 | 49.38 | 2457516.4663 ± 0.0002 | 0.00059 | 0.00053 |

### Table 12. Parameters calculated for HAT-P-22 b

| NO. | $R_P(R_J)$ | K | $T_{14}$(day) | i(°) | BIC | $T_c$ (BJD$_{TDB}$) | $\sigma_{T_{FWHM}}$ | $\sigma_{depth}$ |
|---|---|---|---|---|---|---|---|---|
| 1 | 0.94 ± 0.06 | 0.093 ± 0.001 | 0.1164 ± 0.0004 | 88.5 | 57.18 | 2455614.4272 ± 0.0001 | 0.00026 | 0.00032 |
| 2 | 1.05 ± 0.06 | 0.103 ± 0.001 | 0.1201 ± 0.0002 | 90.0 | 50.48 | 2455630.4828 ± 0.0002 | 0.00024 | 0.00029 |
| 3 | 1.08 ± 0.07 | 0.107 ± 0.002 | 0.1211 ± 0.0005 | 88.4 | 44.67 | 2455948.4979 ± 0.0001 | 0.00033 | 0.00052 |
| 4 | 1.05 ± 0.06 | 0.104 ± 0.001 | 0.1164 ± 0.0005 | 88.8 | 65.26 | 2456327.5369 ± 0.0001 | 0.00030 | 0.00026 |
| 5 | 1.12 ± 0.05 | 0.111 ± 0.001 | 0.1221 ± 0.0002 | 85.8 | 59.55 | 2456356.4504 ± 0.0006 | 0.00013 | 0.00021 |
| 6 | 1.02 ± 0.05 | 0.101 ± 0.001 | 0.1161 ± 0.0002 | 88.5 | 65.78 | 2456687.3126 ± 0.0001 | 0.00014 | 0.00019 |
| 7 | 1.07 ± 0.06 | 0.106 ± 0.001 | 0.1142 ± 0.0002 | 89.1 | 55.83 | 2456780.4644 ± 0.0002 | 0.00022 | 0.00037 |
| 8 | 1.08 ± 0.29 | 0.107 ± 0.024 | 0.0906 ± 0.0062 | 88.8 | 51.40 | 2457024.5856 ± 0.0018 | 0.00362 | 0.00513 |
| 9 | 1.16 ± 0.20 | 0.115 ± 0.015 | 0.1278± 0.0037 | 88.9 | 57.19 | 2457024.5971 ± 0.0010 | 0.00218 | 0.00364 |
| 10 | 1.13 ± 0.05 | 0.112 ± 0.001 | 0.1202 ± 0.0002 | 88.7 | 73.29 | 2457734.5013 ± 0.0002 | 0.00011 | 0.00015 |
| 11 | 1.08 ± 0.07 | 0.107 ± 0.002 | 0.1156 ± 0.0005 | 88.8 | 46.41 | 2457779.4723 ± 0.0001 | 0.00030 | 0.00049 |

### Table 13. Parameters calculated for HAT-P-28 b

| NO. | $R_P(R_J)$ | k | $T_{14}$(day) | i(°) | BIC | $T_c$ (BJD$_{TDB}$) | $\sigma_{T_{FWHM}}$ | $\sigma_{depth}$ |
|---|---|---|---|---|---|---|---|---|
| 1 | 1.42 ± 0.09 | 0.130 ± 0.002 | 0.138 ± 0.004 | 87.3 | 62.50 | 2455808.4618 ± 0.0012 | 0.0024 | 0.00059 |
| 2 | 1.16 ± 0.11 | 0.109 ± 0.003 | 0.101 ± 0.005 | 90.0 | 60.28 | 2455808.4753 ± 0.0015 | 0.0030 | 0.00084 |
| 3 | 1.43 ± 0.09 | 0.134 ± 0.002 | 0.136 ± 0.004 | 90.0 | 45.85 | 2455808.4718 ± 0.0013 | 0.0026 | 0.00076 |
| 4 | 1.34 ± 0.08 | 0.126 ± 0.001 | 0.115 ± 0.002 | 89.4 | 47.28 | 2455808.4717 ± 0.0007 | 0.0014 | 0.00045 |
| 5 | 1.28 ± 0.09 | 0.118 ± 0.002 | 0.150 ± 0.004 | 89.5 | 41.59 | 2455834.5225 ± 0.0012 | 0.0025 | 0.00065 |
| 6 | 1.36 ± 0.09 | 0.126 ± 0.003 | 0.130 ± 0.006 | 89.9 | 43.82 | 2455834.5199 ± 0.0017 | 0.0034 | 0.00076 |
| 7 | 1.18 ± 0.09 | 0.109 ± 0.002 | 0.127 ± 0.001 | 88.4 | 56.21 | 2455834.5251 ± 0.0021 | 0.0022 | 0.00052 |
| 8 | 1.11 ± 0.10 | 0.104 ± 0.002 | 0.121 ± 0.004 | 88.4 | 55.12 | 2455857.3183 ± 0.0012 | 0.0025 | 0.00053 |
| 9 | 1.20 ± 0.08 | 0.114 ± 0.001 | 0.137 ± 0.003 | 88.2 | 55.76 | 2455912.6913 ± 0.0009 | 0.0019 | 0.00039 |
| 10 | 1.19 ± 0.08 | 0.109 ± 0.001 | 0.132 ± 0.003 | 88.4 | 54.03 | 2456508.7671 ± 0.0009 | 0.0019 | 0.00040 |
| 11 | 1.20 ± 0.09 | 0.109 ± 0.002 | 0.127 ± 0.003 | 88.9 | 57.10 | 2456609.7361 ± 0.0009 | 0.0019 | 0.00046 |
| 12 | 1.32 ± 0.09 | 0.122 ± 0.001 | 0.136 ± 0.004 | 88.5 | 59.00 | 2456658.5958 ± 0.0013 | 0.0026 | 0.00048 |
| 13 | 1.06 ± 0.09 | 0.100 ± 0.001 | 0.131 ± 0.003 | 89.1 | 49.27 | 2456886.6009 ± 0.0009 | 0.0018 | 0.00039 |
| 14 | 1.17 ± 0.08 | 0.109 ± 0.001 | 0.128 ± 0.003 | 89.8 | 57.69 | 2456902.8861 ± 0.0004 | 0.0015 | 0.00028 |
| 15 | 1.61 ± 0.08 | 0.151 ± 0.000 | 0.131 ± 0.002 | 89.9 | 53.28 | 2457010.3742 ± 0.0005 | 0.0011 | 0.00025 |
| 16 | 1.23 ± 0.10 | 0.114 ± 0.003 | 0.134 ± 0.006 | 87.4 | 46.35 | 2457326.3225 ± 0.0017 | 0.0034 | 0.00076 |
| 17 | 1.21 ± 0.11 | 0.114 ± 0.004 | 0.130 ± 0.007 | 88.3 | 52.83 | 2457684.6190 ± 0.0022 | 0.0044 | 0.00100 |
| 18 | 1.26 ± 0.08 | 0.118 ± 0.000 | 0.136 ± 0.001 | 89.9 | 47.00 | 2457710.6781 ± 0.0002 | 0.00049 | 0.00017 |
| 19 | 1.38 ± 0.10 | 0.130 ± 0.003 | 0.138 ± 0.007 | 86.2 | 54.56 | 2458404.4685 ± 0.0020 | 0.0041 | 0.00102 |
| 20 | 1.25 ± 0.09 | 0.118 ± 0.002 | 0.134 ± 0.004 | 88.6 | 49.70 | 2458479.3831 ± 0.0012 | 0.0024 | 0.00059 |

### Table 14. Parameters calculated for HAT-P-30 b

| NO. | $R_P(R_J)$ | k | $T_{14}$(day) | i(°) | BIC | $T_c$ (BJD$_{TDB}$) | $\sigma_{T_{FWHM}}$ | $\sigma_{depth}$ |
|---|---|---|---|---|---|---|---|---|
| 1 | 1.43 ± 0.12 | 0.109 ± 0.002 | 0.080 ± 0.004 | 85.1 | 41.74 | 2455650.3980 ± 0.0011 | 0.0023 | 0.00061 |
| 2 | 1.24 ± 0.12 | 0.094 ± 0.002 | 0.070 ± 0.004 | 89.6 | 57.20 | 2455894.9223 ± 0.0012 | 0.0025 | 0.00046 |
| 3 | 1.32 ± 0.12 | 0.100 ± 0.002 | 0.079 ± 0.003 | 88.5 | 49.35 | 2455945.5112 ± 0.0008 | 0.0017 | 0.00044 |
| 4 | 1.40 ± 0.32 | 0.104 ± 0.024 | 0.082 ± 0.007 | 88.6 | 59.30 | 2455970.8145 ± 0.0021 | 0.0042 | 0.00513 |
| 5 | 1.18 ± 0.11 | 0.089 ± 0.001 | 0.081 ± 0.009 | 86.2 | 52.88 | 2455976.4272 ± 0.0026 | 0.0052 | 0.00034 |
| 6 | 1.41 ± 0.12 | 0.109 ± 0.002 | 0.084 ± 0.004 | 83.4 | 41.59 | 2455976.4299 ± 0.0012 | 0.0025 | 0.00055 |
| 7 | 1.36 ± 0.11 | 0.104 ± 0.002 | 0.061 ± 0.003 | 88.6 | 57.78 | 2456001.7266 ± 0.0008 | 0.0015 | 0.00045 |
| 8 | 1.48 ± 0.19 | 0.114 ± 0.010 | 0.126 ± 0.008 | 79.6 | 48.04 | 2456240.6226 ± 0.0024 | 0.0049 | 0.0023 |
| 9 | 1.26 ± 0.13 | 0.094 ± 0.003 | 0.071 ± 0.003 | 89.6 | 57.56 | 2456296.8374 ± 0.0008 | 0.0017 | 0.00061 |
| 10 | 1.24 ± 0.12 | 0.094 ± 0.002 | 0.086 ± 0.004 | 88.3 | 49.63 | 2456687.5067 ± 0.0012 | 0.0024 | 0.00055 |
| 11 | 1.29 ± 0.11 | 0.100 ± 0.001 | 0.090 ± 0.003 | 85.0 | 43.57 | 2456704.3739 ± 0.0009 | 0.0019 | 0.00036 |
| 12 | 1.37 ± 0.27 | 0.104 ± 0.018 | 0.097 ± 0.004 | 87.1 | 57.16 | 2456735.2853 ± 0.0012 | 0.0024 | 0.0039 |
| 13 | 1.33 ± 0.11 | 0.104 ± 0.001 | 0.057 ± 0.003 | 89.6 | 57.00 | 2456982.6255 ± 0.0007 | 0.0016 | 0.00039 |
| 14 | 1.20 ± 0.11 | 0.089 ± 0.001 | 0.076 ± 0.003 | 86.8 | 52.22 | 2457013.5430 ± 0.0010 | 0.0021 | 0.00034 |
| 15 | 1.46 ± 0.11 | 0.114 ± 0.001 | 0.092 ± 0.003 | 83.6 | 35.83 | 2458140.5902 ± 0.0008 | 0.0017 | 0.00039 |
| 16 | 1.45 ± 0.11 | 0.114 ± 0.001 | 0.092 ± 0.002 | 83.1 | 41.90 | 2458528.4551 ± 0.0007 | 0.0015 | 0.00035 |

**Conclusion**

In order to check the light curves of the six planets for which an initial analysis was made using ETD, we examined an independent software package (AIJ) and denoted the reason for using it in the Method section. In the Data Analysis section, we studied a total of 82 light curves and extracted parameters from them such as $R_p$, k, $t_{14}$, $T_c$ and i and have tabulated them along with their uncertainties in Tables 9-14. In order to analyze the light curves recorded in the same passband, we plotted them in units of relative flux with respect to the orbital phase for each of the passbands. Then we analyzed the data with the same filter using AIJ (Table 16). The light curves related to special filters have mentioned (Appendix B.). Then we analyzed the data with the same filter using AIJ. The Precise values of the parameters show that using AIJ as a fitting tool for follow-up observations can lead to results comparable to the NEA (Table 15). The similarities in the parameter values show the merit in analysis of amateur data by making use of the right software packages such as AstroImageJ and web services as EXOFAST, considering the differences in the qualities of photometric data acquired by amateur observers and the precise follow-up light curves from relatively larger telescopes to derive the parameters given in the literature. They also prove the follow-up potential of amateur and university observatories in providing system parameters as well as mid-transit times. Such information will be invaluable considering the numbers of future discoveries from the ground and space-based exoplanet surveys.

**Table 15. Specifications of planets and their host stars (NASA Exoplanet Archive)**

| Planets | $R_p(R_J)$ | k | $t_{14}$(day) | i (deg) | $T_c$ (BJD$_{TDB}$) | R* (R$_{sun}$) | L* (log$_{10}$(L$_{sun}$)) | M* (M$_{sun}$) |
|---|---|---|---|---|---|---|---|---|
| HAT-P-8 b | 1.40 ± 0.13 (Stassun et al. 2017) | 0.0953 ± 0.0009 (Latham et al. 2009) | 0.1587 ± 0.0011 (Latham et al. 2009) | 87.8 ± 0.76 (Stassun et al. 2017) | 2454437.6749 ± 0.0014 (Bonomo et al. 2017) | 1.57 ± 0.15 (Stassun et al. 2017) | 0.5190 ± 0.050 (Latham et al. 2009) | 1.27 ± 0.41 (Stassun et al. 2017) |
| HAT-P-16 b | 1.289 ± 0.066 (Buchhave et al. 2010) | 0.1071 ± 0.0014 (Buchhave et al. 2010) | 0.1276 ± 0.0013 (Buchhave et al. 2010) | 86.6 ± 0.7 (Buchhave et al. 2010) | 2455027.59293± 0.00031 (Buchhave et al.2010) | 1.24 ± 0.054 (Buchhave et al. 2010) | $0.294^{+0.046}_{-0.051}$ (Buchhave et al. 2010) | 1.22 ± 0.039 (Buchhave et al. 2010) |
| HAT-P-21 b | 1.11 ± 0.16 (Stassun et al. 2017) | 0.0950 ± 0.0022 (Bakos et al. 2011) | 0.1530 ± 0.0027 (Bakos et al. 2011) | 87.2 ± 0.70 (Stassun et al. 2017) | 2454996.41312± 0.00069 (Bakos et al. 2011) | 1.21 ± 0.17 (Stassun et al. 2017) | $0.1481^{+0.007184}_{-0.007304}$ (Brown et al. 2018) | 1.24 ± 0.60 (Stassun et al. 2017) |
| HAT-P-22 b | 1.15 ± 0.19 (Stassun et al. 2017) | 0.1065 ± 0.0017 (Turner et al. 2016) | 0.1196 ± 0.0014 (Bakos et al. 2011) | 86.9 ± 0.55 (Stassun et al. 2017) | 2454930.22001± 0.00025 (Bakos et al. 2011) | 1.11 ± 0.05 (Stassun et al. 2017) | $-0.114^{+0.048}_{-0.054}$ (Bakos et al. 2011) | 1.13 ± 0.22 (Stassun et al. 2017) |
| HAT-P-28 b | $1.212^{+0.113}_{-0.082}$ (Buchhave et al. 2011) | 0.1130 ± 0.0024 (Buchhave et al. 2011) | 0.1341 ± 0.0020 (Buchhave et al. 2011) | 88.0 ± 0.9 (Buchhave et al. 2011) | 2455417.59832± 0.00053 (Buchhave et al. 2011) | $1.10^{+0.091}_{-0.069}$ (Buchhave et al. 2011) | $0.053^{+0.080}_{-0.066}$ (Buchhave et al. 2011) | 1.03 ± 0.047 (Buchhave et al. 2011) |
| HAT-P-30 b | 1.44 ± 0.15 (Stassun et al. 2017) | $0.1109^{+0.0016}_{-0.0014}$ (Maciejewski et al. 2016) | 0.0887 ± 0.0015 (Johnson et al. 2017) | 83.6 ± 0.04 (Stassun et al. 2017) | 2455456.46561± 0.00037 (Johnson et al. 2011) | 1.31 ± 0.13 (Stassun et al. 2017) | $0.4097^{+0.00615}_{-0.00623}$ (Brown et al. 2018) | 1.55 ± 0.51 (Stassun et al. 2017) |

**Table 16. AIJ's results with the same filter**

| Planets | $R_p(R_J)$ | k | $t_{14}$(day) | i (deg) |
|---|---|---|---|---|
| HAT-P-8 b (R) | 1.41 ± 0.12 | 0.0943 ± 0.0019 | 0.1653 ± 0.0012 | 89.4 |
| HAT-P-8 b (Clear) | 1.46 ± 0.11 | 0.0959 ± 0.0017 | 0.1634 ± 0.0035 | 89.9 |
| HAT-P-16 b (R) | 1.29 ± 0.05 | 0.1077 ± 0.0014 | 0.1251 ± 0.0010 | 89.6 |
| HAT-P-16 b (V) | 1.33 ± 0.05 | 0.1109 ± 0.0017 | 0.1271 ± 0.0009 | 89.8 |
| HAT-P-16 b (I) | 1.20 ± 0.06 | 0.0994 ± 0.0018 | 0.1326 ± 0.0006 | 89.3 |
| HAT-P-16b (Clear) | 1.31 ± 0.06 | 0.1224 ± 0.0029 | 0.1343 ± 0.0023 | 88.6 |
| HAT-P-21 b (R) | 1.17 ± 0.16 | 0.0994 ± 0.0023 | 0.1559 ± 0.0004 | 89.7 |
| HAT-P-21 b (Clear) | 1.11 ± 0.16 | 0.0948 ± 0.0023 | 0.1484 ± 0.0007 | 89.9 |
| HAT-P-22 b (R) | 1.16 ± 0.06 | 0.1072 ± 0.0017 | 0.1194 ± 0.0007 | 89.7 |
| HAT-P-22 b (V) | 1.10 ± 0.06 | 0.1019 ± 0.0021 | 0.1175 ± 0.0005 | 89.5 |
| HAT-P-22 b (Clear) | 1.11 ± 0.06 | 0.1034 ± 0.0015 | 0.1149 ± 0.0002 | 89.4 |
| HAT-P-28 b (R) | $1.25 \pm 0.09$ | $0.1170 \pm 0.0022$ | $0.1361 \pm 0.0037$ | 89.9 |
| HAT-P-28 b (Clear) | $1.25 \pm 0.09$ | $0.1178 \pm 0.0027$ | $0.1296 \pm 0.0032$ | 89.9 |
| HAT-P-30 b (R) | $1.41 \pm 0.27$ | $0.1113 \pm 0.0186$ | $0.0868 \pm 0.0041$ | 87.5 |
| HAT-P-30 b (V) | $1.31 \pm 0.29$ | $0.1029 \pm 0.0195$ | $0.0888 \pm 0.0038$ | 89.9 |
| HAT-P-30 b (I) | $1.25 \pm 0.28$ | $0.0979 \pm 0.0180$ | $0.0899 \pm 0.0053$ | 89.6 |
| HAT-P-30 b (Clear) | $1.38 \pm 0.26$ | $0.1063 \pm 0.0175$ | $0.0916 \pm 0.0031$ | 87.3 |

## Acknowledgements

This manuscript was prepared by the International Occultation Timing Association Middle East section (IOTA/ME). We would like to thank the local and scientific organizers of the IOTA/ME Photometric Observations of Exoplanet Transits Workshop held in Ankara University between 13-18 October 2018 for the hands-on exercises in exoplanet transit analyses.

## Appendix A.

In figures A1-A6, the horizontal axis is plotted based on the mid-point of the transit and the zero point in the graph is equal to $T_c$ based on BJD. The vertical axis is based on delta magnitude of the star and the zero point is the mean of average of delta magnitude.

The pink color signifies that the light curve was made using an I filter; the red color denotes an R filter; blue color denotes the Clear filter, and the green color, the V filter. Each graph is marked with a number according to the Tables 2-7.

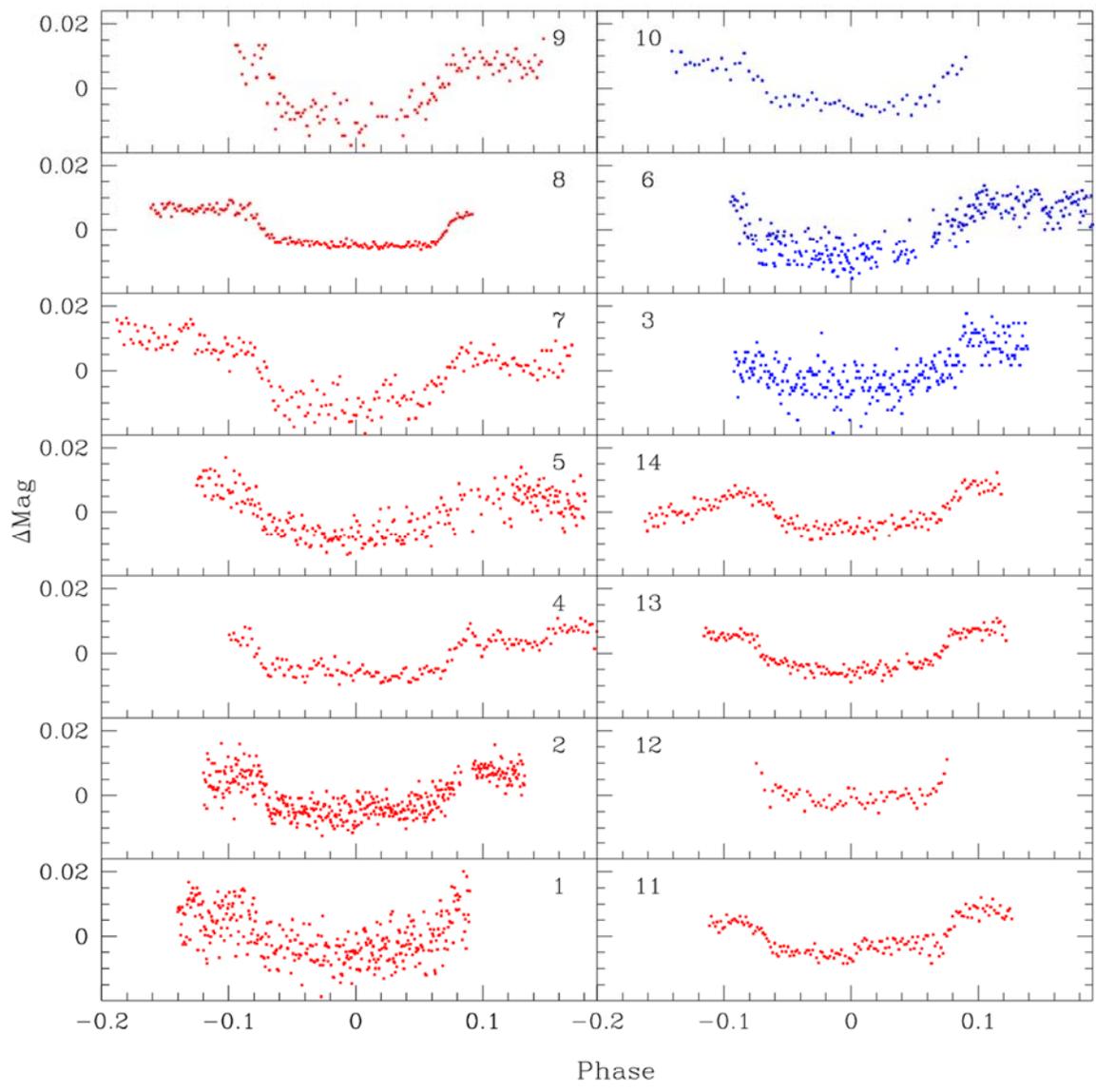

**A1. Light curves of Planet HAT-P-8 b**

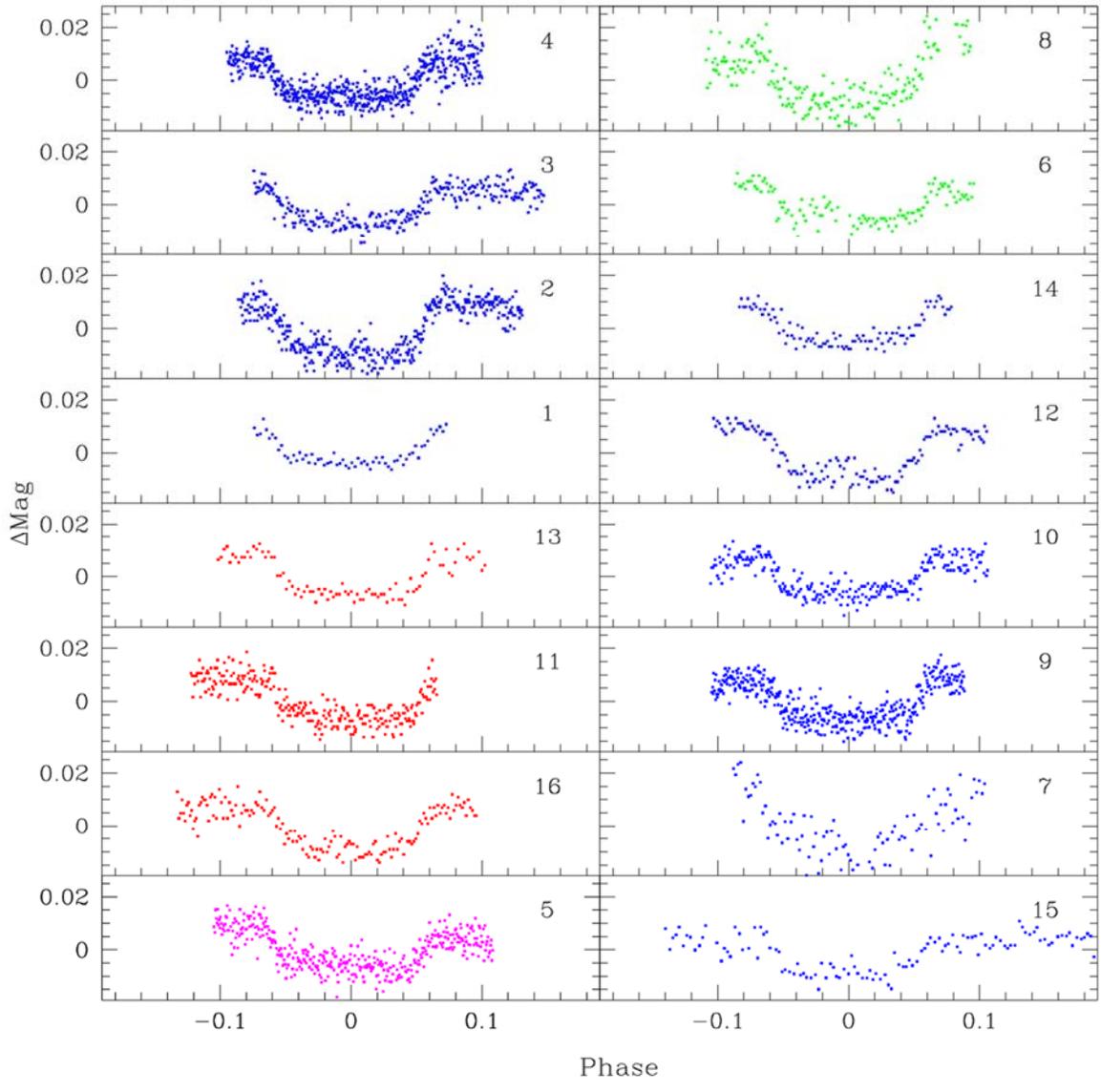

**A2. Light curves of Planet HAT-P-16 b**

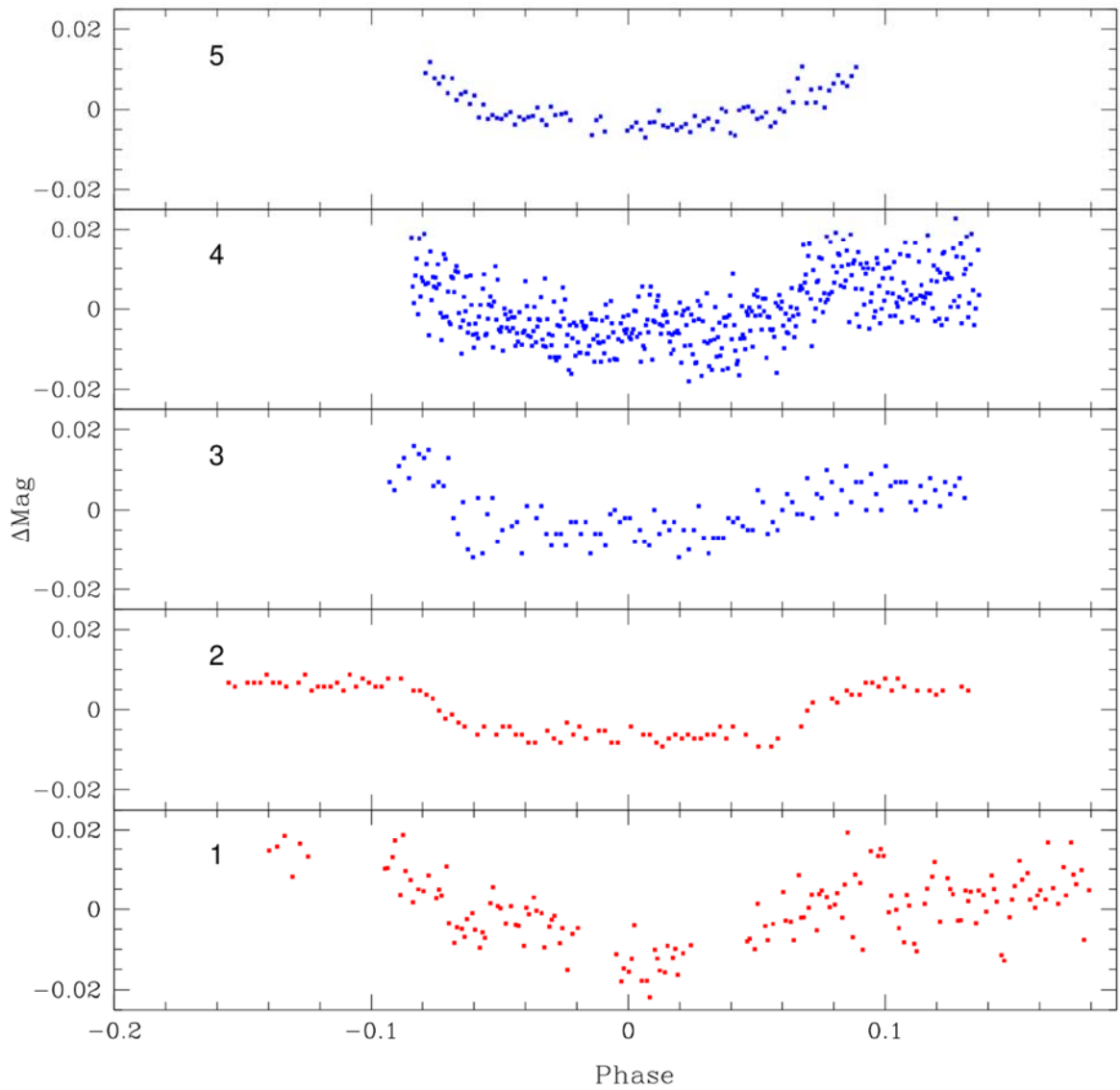

**A3. Light curves of Planet HAT-P-21 b**

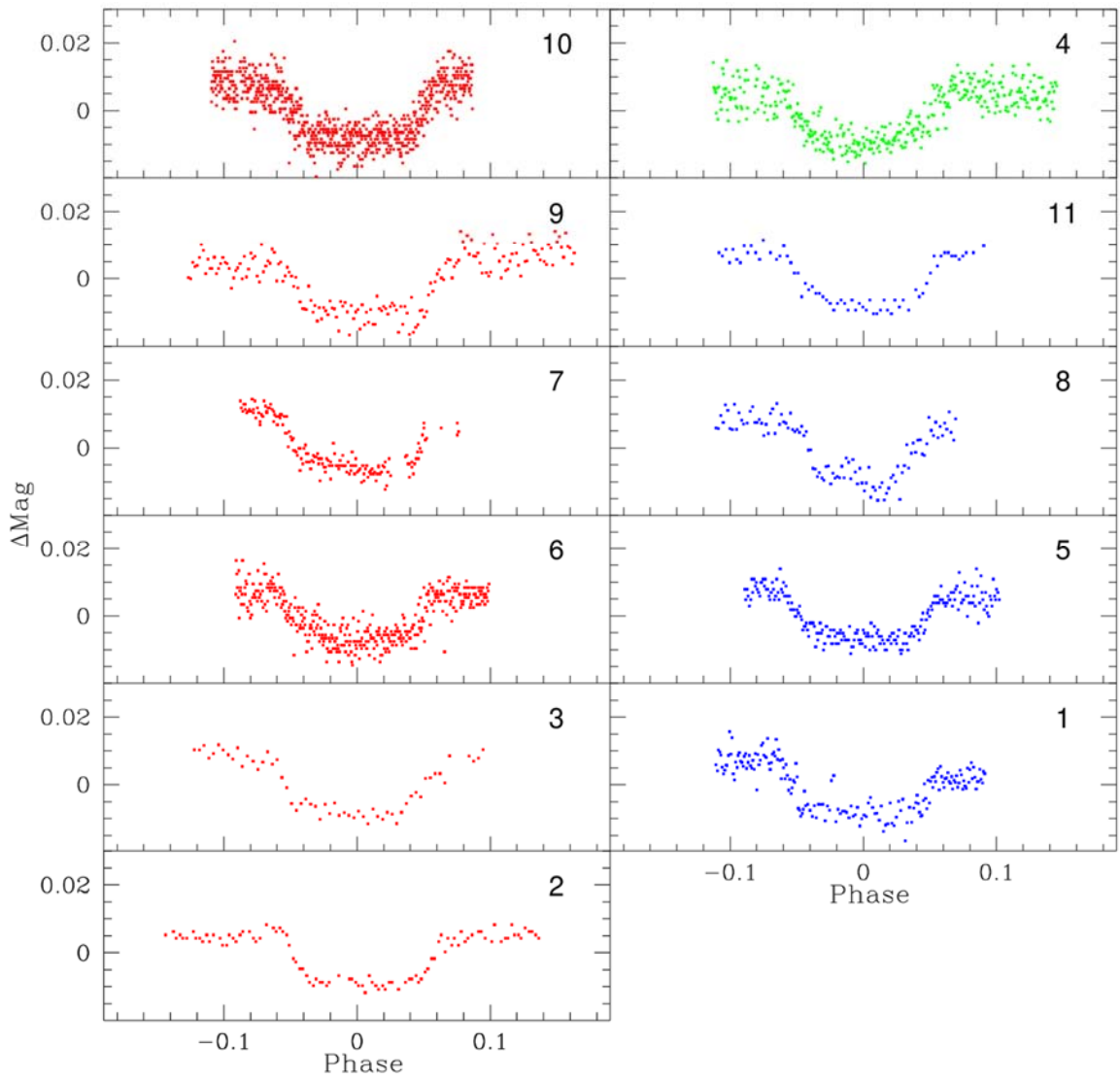

**A4. Light curves of Planet HAT-P-22 b**

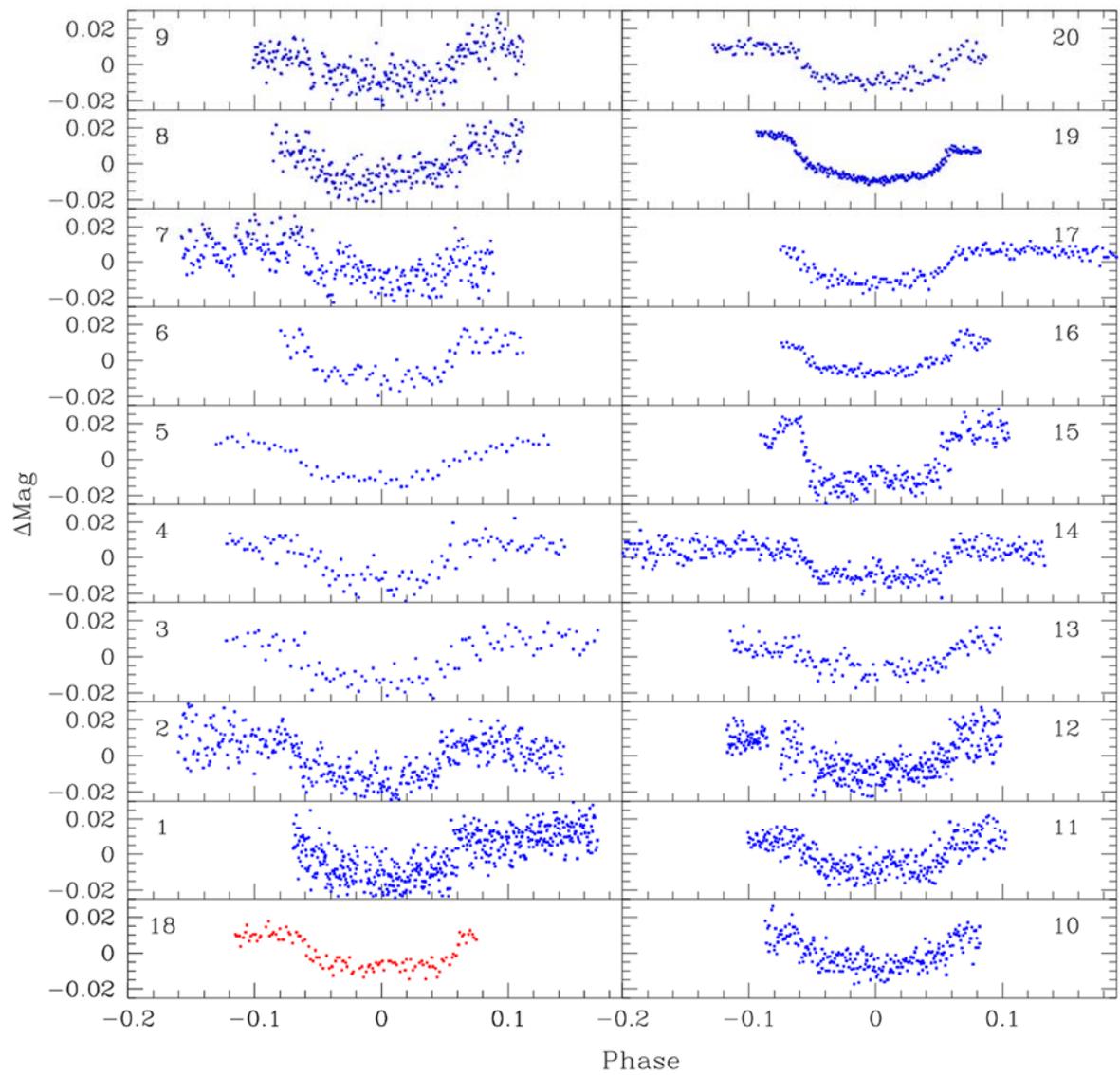

**A5. Light curves of Planet HAT-P-28 b**

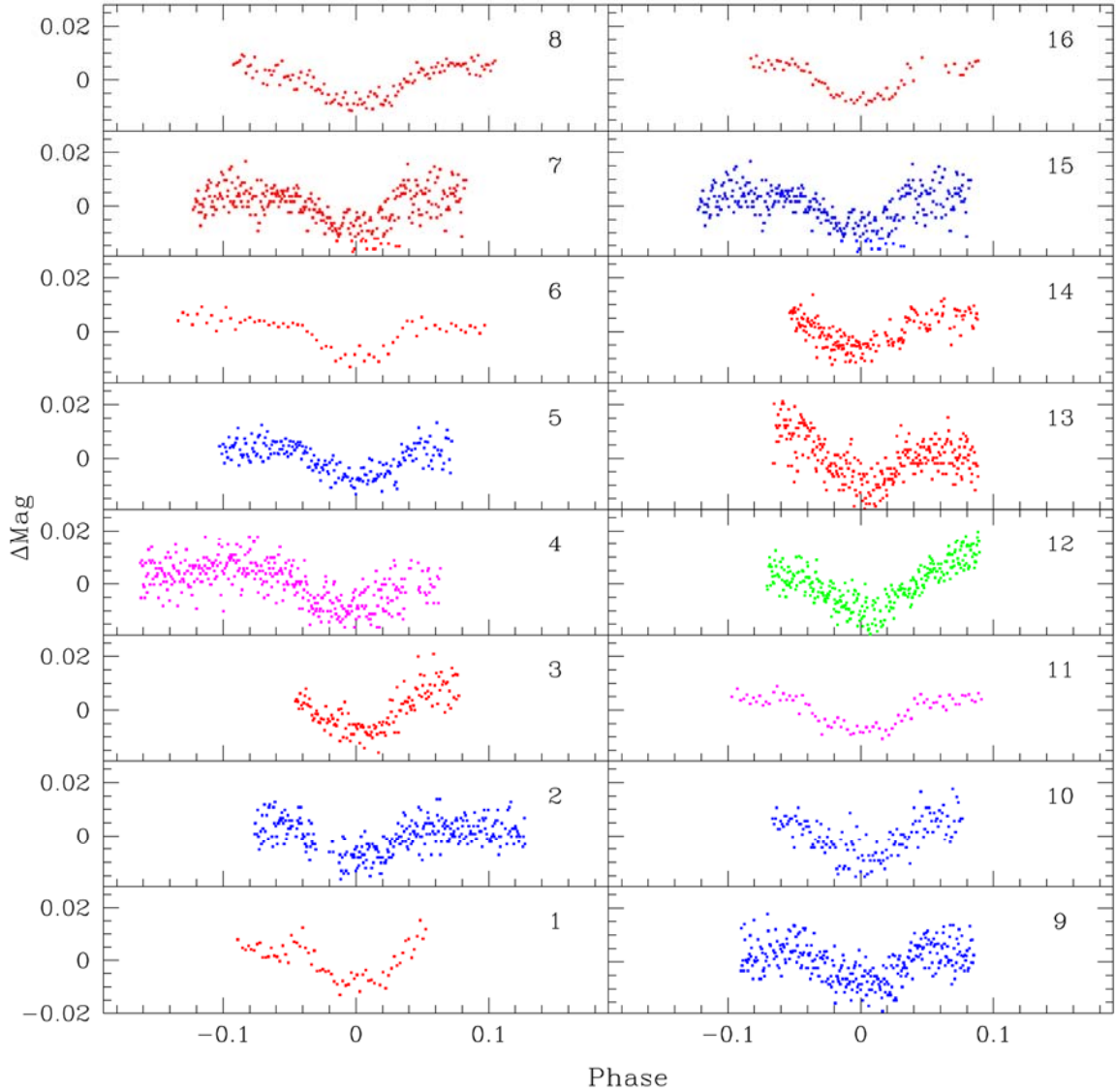

**A6. Light curves of Planet HAT-P-30 b**

## Appendix B.

In figures B1-B6, the horizontal axis is plotted based on Phase and the zero point in the graph is equal to $T_c$ (BJD$_{TDB}$). The vertical axis is based on the flux of the star that is normalized to 1. The raw data are shown as solid blue dots which are plotted with no fit, while the solid magenta dots show the same data assuming that it will reflect a dip in brightness due to the exoplanet's transit, which is shown by the magenta line through the data. The duration of the transit completely lies inside the predicted times of ingress and egress.

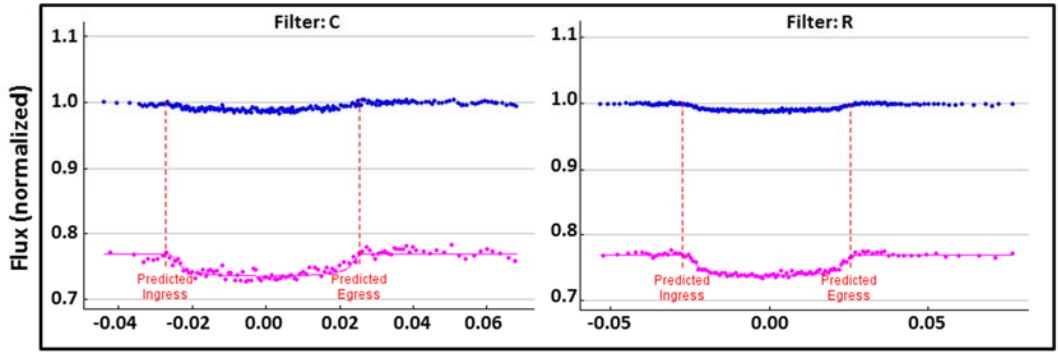

**B1. HAT-P-8 b**

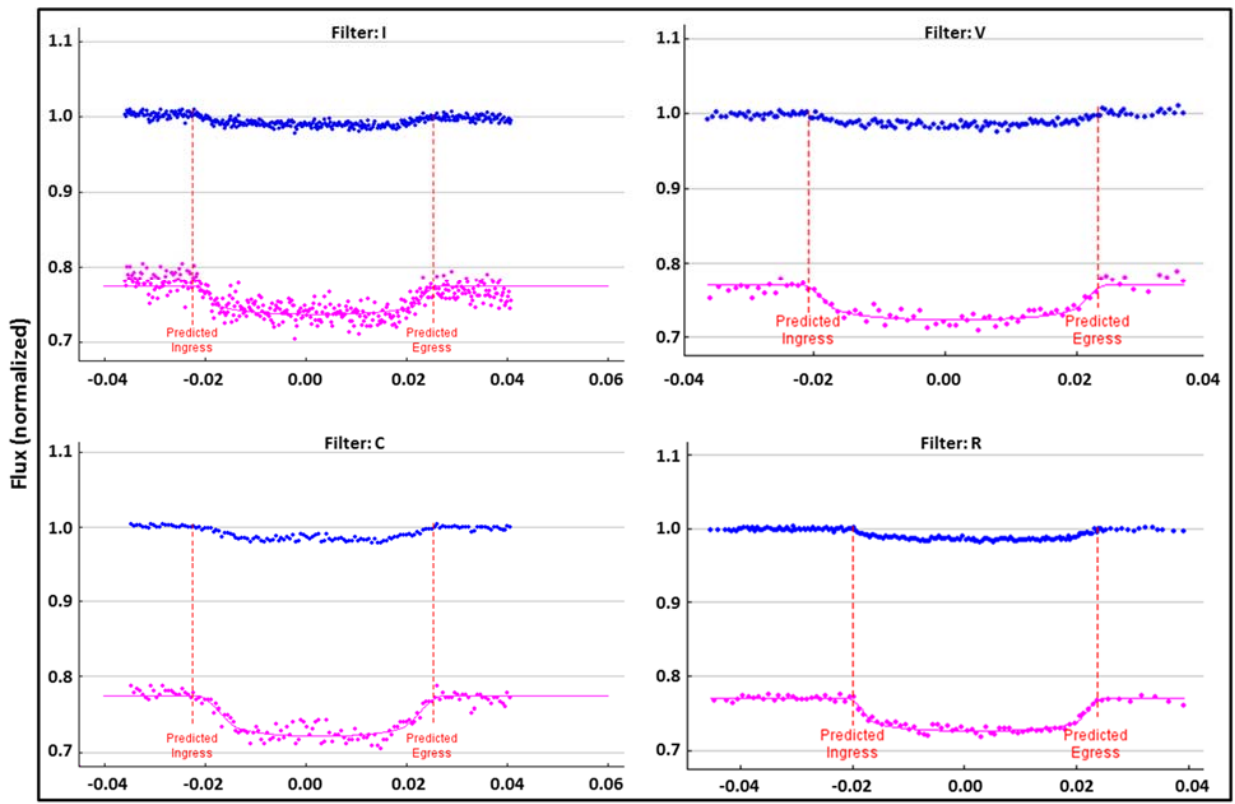

**B2. HAT-P-16 b**

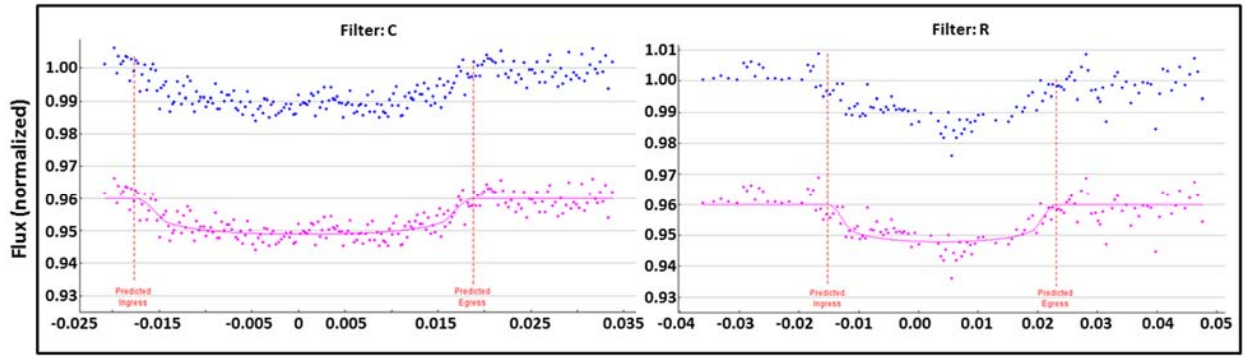

**B3. HAT-P-21 b**

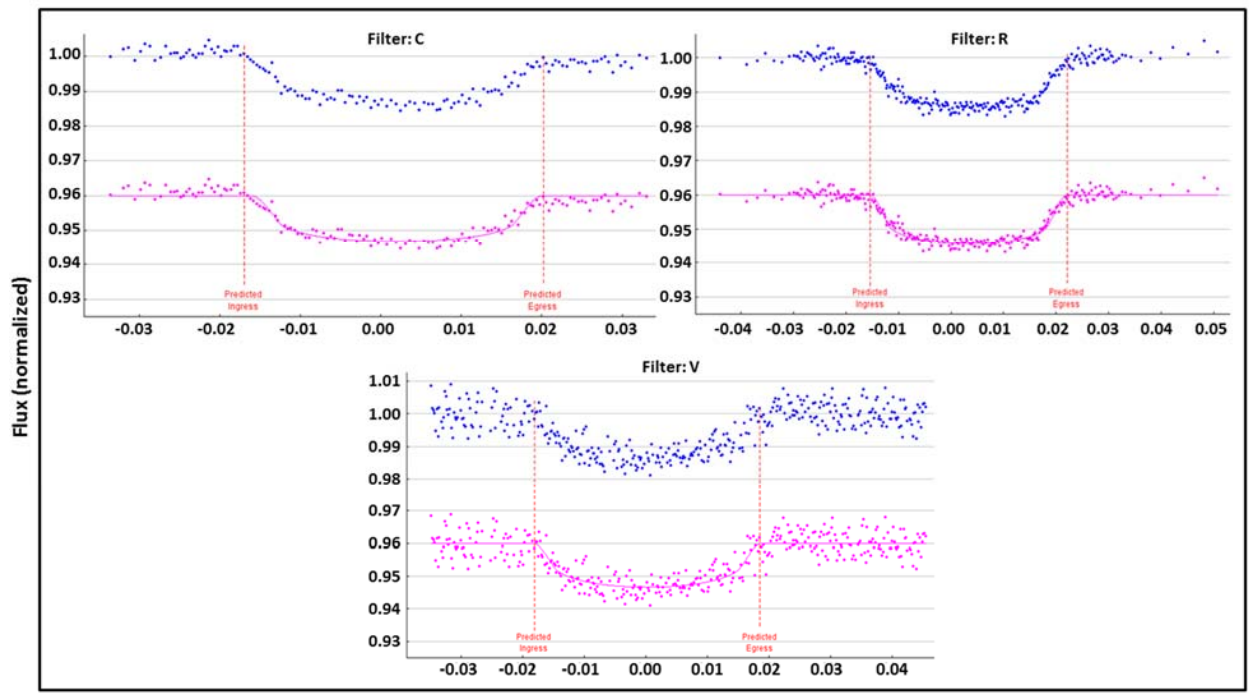

**B4. HAT-P-22 b**

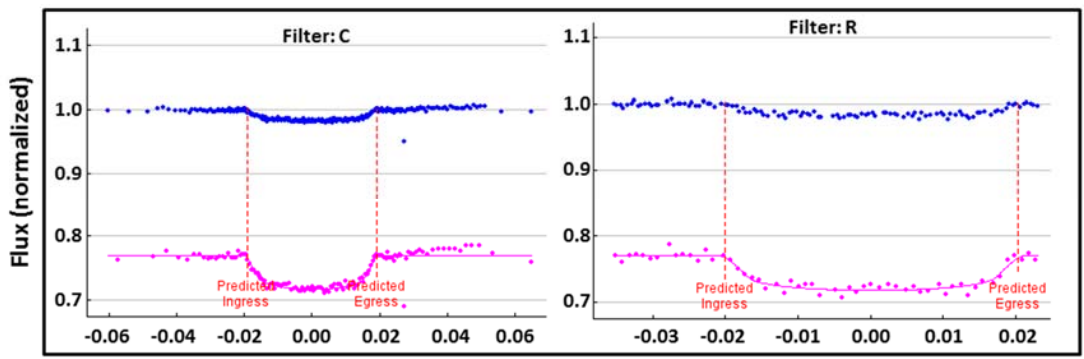

**B5. HAT-P-28 b**

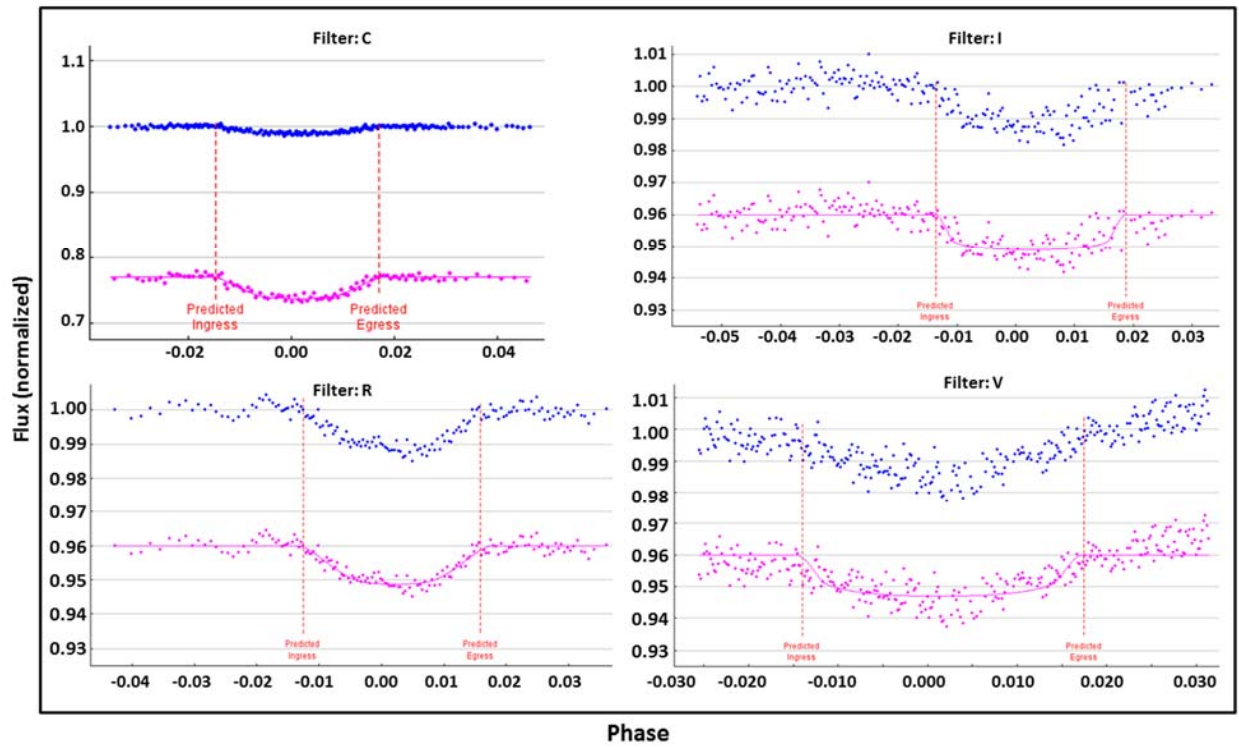

**B6. HAT-P-30 b**